\documentclass[%
 reprint,
 amsmath,amssymb,
 aps,
]{revtex4-1}

\usepackage{graphicx}
\usepackage{dcolumn}
\usepackage{bm}
\usepackage{color}
\usepackage[colorlinks=true, linkcolor=blue, citecolor=blue, urlcolor=blue]{hyperref}

\def\ldef{\mathrel{\mathop:}=}
\def\e{\mathrm{e}}
\def\z{z}

\def\d{\mathrm{d}}
\def\n{\mathrm{n}}
\def\Re{\mathrm{Re}}
\def\Im{\mathrm{Im}}
\def\mysec#1{{\bf #1 --}}

\begin{document}

\title{Exact Quantum Trace Formula from Complex Periodic Orbits}

\author{Chaoming Song}%
 \email{c.song@miami.edu}
\affiliation{%
Department of Physics, University of Miami, Coral Gables FL, 33146 USA.
}%

\date{\today}


\begin{abstract}
The Gutzwiller trace formula establishes a profound connection between the quantum spectrum and classical periodic orbits. However, its application is limited by its reliance on the semiclassical saddle point approximation. In this work, we explore the full quantum version of the trace formula using the Lefschetz thimble method by incorporating complexified periodic orbits. Upon complexification, classical real periodic orbits are transformed into cycles on compact Riemann surfaces. Our key innovation lies in the simultaneous complexification of the periods of cycles, resulting in a fully quantum trace formula that accounts for all contributions classified by the homology classes of the associated Riemann surfaces. This formulation connects the quantum spectrum to contributions across all complex time directions, encompassing all relevant homology classes. Our approach naturally unifies and extends two established methodologies: periodic orbits in real time, as in Gutzwiller's original work, and quantum tunneling in imaginary time, as in the instanton method.
\end{abstract}

\maketitle

\mysec{Introduction} 
Understanding the intricate relationship between quantum spectra and classical dynamics is a fundamental pursuit in physics. The celebrated Gutzwiller trace formula \cite{gutzwiller1971periodic,gutzwiller2013chaos} establishes a profound connection between the quantum spectrum and classical periodic orbits, making it a cornerstone for elucidating quantum chaos. For instance, in systems exhibiting geodesic motion on hyperbolic surfaces, the Gutzwiller trace formula coincides with the exact Selberg trace formula \cite{selberg1957harmonic}, owing to the system's high symmetry. However, beyond these specific instances, the Gutzwiller trace formula remains a semiclassical approximation \cite{berry1972semiclassical, berry1977regular, cvitanovic1993fredholm, brizard2017motion, PhysRev.96.1124}, inherently limited to capturing only perturbative aspects of the quantum energy spectrum.

A significant limitation of semiclassical methods is their inability to incorporate nonperturbative effects, such as quantum tunneling, which manifest as exponentially small contributions in weak coupling regimes. Traditionally, these effects are qualitatively described by instantons, i.e., classical solutions in imaginary time \cite{coleman1977fate, voloshin1975bubbles, callan1977fate, Bogomolny1980431, Coleman198802}, that are absent from real-time classical solutions. Despite their importance, a systematic framework for accounting for instanton contributions, particularly their quantum interference \cite{dunne2012resurgence, dunne2014uniform, zinn1981multi, zinn1983multi}, remains challenging.

Recent advancements have turned to Picard-Lefschetz theory as a promising approach to real-time path integrals \cite{witten2010new, witten2011analytic, tanizaki2014real, Cherman:2014sba}. This method can be viewed as an infinite-dimensional extension of contour integration \cite{milnor1963morse, arnold_mechanics}, where the original path integral over real functional space is extended into a complexified domain. By deforming the integration contours into the complex plane, contributions from complex saddle points can be systematically included \cite{Aarts:2013fpa, Cristoforetti:2012su, Cristoforetti:2013wha, Cristoforetti:2014gsa, Fujii:2013sra, Mukherjee:2014hsa}, thereby capturing the full nonperturbative nature of the path integral. In principle, this allows for the exact computation of path integrals, naturally incorporating all nonperturbative effects \cite{Unsal:2012zj, Basar:2013eka, Cherman:2014ofa, Cherman:2014sba}. Although there have been calls to derive an exact quantum trace formula \cite{berry1972semiclassical, berry1977regular,nekrasov2018tying}, previous attempts, such as those rooted in the exact WKB perspective \cite{sueishi2020exact}, have yielded valuable insights but have fallen short of providing a fully explicit trace formulation.

In this paper, we present a fully quantum version of the trace formula using the complexified path integration over the free loop space, as detailed in Ref. \cite{witten2010new}. Our key innovation, in contrast to prior studies, is the simultaneous complexification of the periods of the orbits. This process allows a classical real orbit to be analytically continued to a cycle on a Riemann surface. To construct the full quantum trace formula, we include contributions from periodic orbits spanning all complex time directions, encompassing all relevant homology classes of these Riemann surfaces. Consequently, our approach naturally generalizes both real-time and imaginary-time methods as special cases. The resulting trace formula provides a new framework to capture all nonperturbative effects in quantum systems, potentially offering a systematic approach applicable to nonperturbative quantum field theory.

\mysec{Semiclassical trace formula} 
Consider an \( n \)-dimensional quantum system with a compact Hamiltonian operator \( \hat{H} \). For simplicity, we set $\hbar = 1$ throughout this paper. The density of states \( d(E) \) can be expressed by the inverse Fourier transform
\begin{equation}\label{eq:d0}
d(E) = \frac{1}{2\pi } \int_{-\infty}^\infty dt\, e^{i E t  } \operatorname{Tr}[ \hat{U}(t) ],
\end{equation}
where \( \hat{U}(t) \ldef e^{-i \hat{H} t } \) is the time evolution operator. We assume \( E \) is real-valued. By representing \( \operatorname{Tr}[ \hat{U}(t) ] \) as a path integral over closed trajectories in phase space \( M = (\mathbb{R}^{2n}, \omega) \) equipped with a symplectic form \( \omega \ldef \sum_{i} dq_i \wedge dp_i \), we have
\[
\operatorname{Tr}[ \hat{U}(t) ] = \int_{\mathcal{L}_t M} \mathcal{D}[ q ] \mathcal{D}[ p ]\, e^{i S[z]  },
\]
where \( \mathcal{L}_t M \) denotes the loop space of phase-space trajectories \( z(t) = (q(t), p(t)) \) with period \( t \) (i.e., \( z(0) = z(t) \)), and the action functional is \( S[ z ] = \int_0^t \left( p \cdot \dot{q} - H( p, q ) \right) dt \). By substituting the path integral for \( \operatorname{Tr}[ \hat{U}(t) ] \) into Eq. \eqref{eq:d0}, we obtain
\begin{equation}\label{eq:d}
d(E) = \frac{1}{2\pi } \int_{-\infty}^\infty \d T \int_{\mathcal{L}_T M} \mathcal{D}[q]\mathcal{D}[p] \, e^{i \mathcal{S}[z]  },
\end{equation}
where the reduced action is defined as \( \mathcal{S}[z] \ldef E T + S[z] = \oint p \, dq + \int_0^T (E - H(q, p)) \, dt \). Here, the period \( T \) allows for negative values. 

To evaluate \( d(E) \), we perform a stationary phase approximation of the path integral. The critical points of \( \mathcal{S}[z] \) correspond to classical periodic orbits. Applying the saddle-point approximation to Eq.~\eqref{eq:d}, we obtain the Gutzwiller semiclassical trace formula:
\begin{equation} \label{eq:sctrace}
d(E) \approx \Gamma(E) +  \sum_{p} \sum_{r=1}^\infty A_{p,r} \, e^{i r \left({\mathcal{S}_p} - \frac{\pi \mu_p}{2}\right)}, 
\end{equation}
where the volume of the energy surface 
\begin{equation}
    \Gamma(E) \ldef  \frac{1}{(2\pi )^n} \int_{\mathbb{R}^{2n}} \d q \, \d p \, \delta(E - H(p, q)),
\end{equation}
accounts for contributions from short-time, non-periodic paths. The second term in Eq. \eqref{eq:sctrace} arises from contributions of classical periodic orbits. Here,
\( p \) indexes the primitive periodic orbits, and \( r \) is the repetition number. The Maslov index \( \mu_p \) is a topological invariant associated with the orbit \( p \), counting the number of sign changes in the eigenvalues of the monodromy matrix \( \mathbf{M}_p \) along the orbit’s trajectory, representing phase shifts due to caustics. The coefficients \( A_{p,r} \ldef \frac{|T_p|}{\pi\sqrt{\left|\det\left(\mathbf{M}_p^r - \mathbf{I}\right)\right|}} \) account for quantum fluctuations around the critical points, with \( T_p \) being the period of orbit \( p \).

\mysec{Lefschetz thimble} Before obtaining the full quantum corrections to the path integral form~\eqref{eq:d}, we first consider an $n$-dimensional real-space integral:
\[
I \ldef \int_{\mathbb{R}^n} \d^n \z \, \exp(f(\z)),
\]
where $\d^n \z \ldef \d z_1 \cdots \d z_n$, and the exponent function $f(\z)$ is assumed to be analytic in a neighborhood of $\z \ldef (z_1, \ldots, z_n) \in \mathbb{C}^n$ and purely imaginary for real $z$.

To evaluate this integral, it is useful to deform the original real integration domain $\mathbb{R}^n$ into complex space, applying Picard-Lefschetz theory. This theory relates the integral to sums over specialized complex contours called thimbles and dual thimbles, which are associated with the critical points of $f(\z)$. By complexifying the variables $z_i$, the original integration domain $\mathbb{R}^n$ becomes a middle-dimensional real submanifold $\mathcal{C}_\mathbb{R}$ in the $2n$-dimensional real manifold $\mathbb{C}^n$.

We treat $\Re\left( f(\z) \right)$ as a Morse function, meaning that it is smooth and its critical points are non-degenerate, with distinct real values. This allows us to use its gradient flow to define the thimbles and dual thimbles. Specifically, a thimble, denoted by $\mathcal{J}_\alpha$, is the set of all points in $\mathbb{C}^n$ that flow toward the critical point $\z_\alpha$ along paths of steepest descent. Similarly, the dual thimble $\mathcal{K}_\alpha$ consists of all points flowing away from $\z_\alpha$ along paths of steepest ascent. These paths satisfy the gradient flow equations:
\[
\frac{\d z_i}{\d \tau} = - g^{i\bar{j}} \overline{\left(\frac{f(z)}{\partial z_{\bar{j}}} \right)},
\]
where $\tau \in \mathbb{R}$ is the flow parameter, $g^{i\overline{j}}$ is the inverse of a Hermitian metric $g_{i\overline{j}}$ on $\mathbb{C}^n$, and the overline denotes complex conjugation. The Lefschez thimble corresponds to the flow from $\tau = -\infty$ whereas the dual thimble corresponds to the flow from $\tau = +\infty$.
Along these flows, the real and imaginary parts of $f(\z)$ evolve as:
\[
\frac{\d}{\d \tau} \Re\, (f) = - \left| \nabla f \right|^2_g, \quad \frac{\d}{\d \tau} \Im\, (f) = 0,
\]
where $\left| \nabla f \right|^2_g \ldef g^{i\overline{j}} \left( \frac{\partial f}{\partial z_i} \right) \overline{\left( \frac{\partial f}{\partial z_j} \right)} \geq 0$. This means that $\Re\, (f)$ decreases along thimbles and increases along dual thimbles, while $\Im\, (f)$ remains constant.

The original integral $I$ can now be expressed as a sum over contributions from the thimbles: \begin{equation}\label{eq:Lefschetz} 
I = \sum_\alpha n_\alpha , \e^{f(\z_\alpha)} \int_{\mathcal{J}_\alpha} \d\mu_{\mathcal{J}_\alpha} \exp\big(\Delta f(\z)\big), 
\end{equation} 
where $\Delta f(\z) \ldef f(\z) - f(\z_\alpha)$, and  $\d\mu_{\mathcal{J}_\alpha}$ is the measure on the thimble $\mathcal{J}_\alpha$ induced from the metric $g$. Since $\Delta f(\z)$ is real and nonpostive along $\mathcal{J}_\alpha$, the integrand $\exp\big(\Delta f(\z)\big)$ is exponentially suppressed away from $\z_\alpha$. This behavior makes the integration over each thimble more manageable compared to integration over the original real domain $\mathbb{R}^n$.

The integer $\n_\alpha \ldef \langle \mathcal{K}_\alpha, \mathcal{C}_\mathbb{R} \rangle \in \mathbb{Z}$ are intersection numbers between the dual thimbles $\mathcal{K}_\alpha$ and the real contour $\mathcal{C}_\mathbb{R}$. Since $\Re\left( f(\z) \right) = 0$ for a real $z$, we have $n_\alpha = 0$ for $\Re\left( f(\z_\alpha) \right) \geq 0$, as there is no upward flow from the critical point $\z_\alpha$ to $\mathcal{C}_\mathbb{R}$, except when the critical point $\z_\alpha$ is real, where a trivial flow exists, making $n_\alpha = 1$. The value of $n_\alpha$ for $\Re\left( f(\z_\alpha) \right) < 0$ is generally unknown and depends on the specifics of the flow equation.

At an isolated nondegenerate critical point $\z_\alpha$, the Hessian of $\Re\left(f(\z)\right)$ has eigenvalues that occur in complex conjugate pairs due to the analyticity of $f$. This structure ensures that the Morse index, defined as the number of negative eigenvalues of the Hessian, is equal to $n$. Consequently, both the thimble $\mathcal{J}_\alpha$ and its dual thimble $\mathcal{K}_\alpha$ are real $n$-dimensional submanifolds, making them middle-dimensional in the real $2n$-dimensional space $\mathbb{C}^n$.

\begin{figure*}[!htb]
    \centering
    \includegraphics[width=\textwidth]{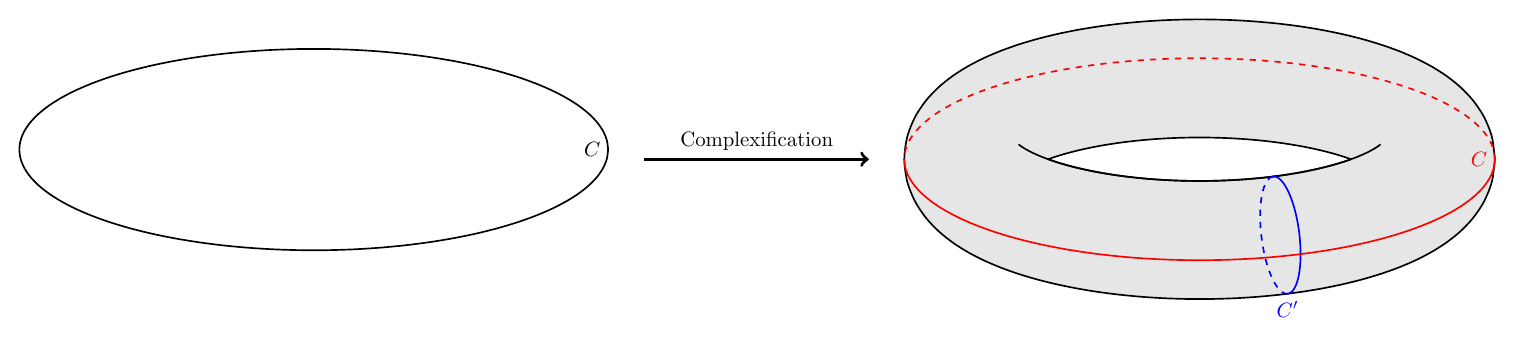}
    \caption{A periodic orbit \((q(t), p(t))\) in the real phase space transforms into a torus after complexification for the double-well potential. Note that the torus represents the underlying Riemann surface associated with both \(q(z)\) and \(p(z)\), rather than the direct hypersurface \((q(z), p(z))\) in the complexified phase space.}
    \label{fig:complex}
\end{figure*}

When critical points are not isolated but instead form submanifolds, referred to as critical submanifolds, the standard definition of Lefschetz thimbles requires modification. Given a critical submanifold $\mathcal{M}_\alpha$ with complex dimension $d_\alpha$, there are $2d_\alpha$ zero modes in the Hessian. This implies that the Morse index is $n-2d_\alpha$. To make the thimbles middle-dimensional, we need to impose $d_\alpha$ additional constraints, effectively splitting the $2d_\alpha$ zero modes evenly between the thimble and the dual thimble. To achieve this, one can select a cycle $\mathcal{M}_\alpha^* \subset \mathcal{M}_\alpha$ with real dimension $d_\alpha$. The corresponding Lefschetz thimbles consist of points that can be reached via downward gradient flow starting from each point on $\mathcal{M}_\alpha^*$.

\mysec{Critical manifolds} To apply Lefschetz thimbles to Eq.~\eqref{eq:d}, we introduce a reparameterization \( \tilde{z}(\eta) \ldef z(\eta T) \) with \( \tilde z(\eta) = \tilde z(\eta+1) \) that factors out the period \( T \), leading to
\begin{align}
  d(E)  = &\frac{1}{2\pi} \int_{-\infty}^\infty \d T \int_{\mathcal{L}_M} \mathcal{D}[\tilde q]\mathcal{D}[\tilde p] \, e^{i \tilde{\mathcal{S}}[\tilde z]  },
\end{align}
with the reparameterized reduced action functional 
$$\tilde{\mathcal{S}}[\tilde z]  \ldef \oint \tilde p \cdot d\tilde q + T\int_0^{1} (E-H(\tilde p,\tilde q)) d\eta.$$ 
Here, \( \mathcal{L}_M \) denotes the loop space of normalized periodic functions mapping \( S^1 \) to \( M \).

Following the discussion from the previous section, we need to deform the original integration domain \(\mathcal{L}_M \times \mathbb{R}\) into a complex space. The real phase space \(M\) is complexified into \(\widehat{M}\). The symplectic form naturally extends to this complexified phase space as \(\widehat{\omega} \ldef \sum_{i} \d q_i \wedge \d p_i\). Additionally, the period \( T \) is also complexified. Therefore, the original integration domain becomes a real integration contour in the ambient space \(\mathcal{L}_{\widehat{M}} \times \mathbb{C}\).

The critical points of \(\delta \tilde{\mathcal{S}} = 0\) correspond to two equations:
\begin{equation} \label{eq:critical}
    \frac{\partial \tilde z}{\partial \eta} = T \Omega \nabla_{\tilde z}H, \quad  H\left( \tilde q, \tilde p\right)  = E,
\end{equation}
where the first equation is Hamilton’s equation, with the symplectic matrix \(\Omega \ldef \begin{pmatrix} 0 & I \\ -I & 0 \end{pmatrix}\), and the second ensures motion restricted to the complexified energy surface \(\widehat \Sigma_E\) parameterized by the energy \(E\), which is generally a complex algebraic variety.

The critical points, as solutions to Eq.~\eqref{eq:critical}, divide naturally into two distinct types: (i) zero-period orbits and (ii) finite-period orbits, corresponding to the smooth and oscillatory contributions to the density of states. We analyze each case below.

(i) \emph{Zero-period orbits}: For zero period $T=0$, Eq.~\eqref{eq:critical} reduces to a constant map, such that $(\tilde q(\eta), \tilde p(\eta)) = (\tilde q_0, \tilde p_0) \in \widehat\Sigma_E$ remains unchanged over time. Consequently, the set of critical points is given by $\mathcal{M}_0 \ldef \widehat \Sigma_E \times \{0\}$, which corresponds to the complex energy surface with a complex dimension $2n-1$. The reduced action $\tilde{\mathcal{S}} = 0$ holds for all critical points within $\mathcal{M}_0$. A natural choice for a subset $\mathcal{M}_0^*$ with real dimension $2n-1$ is to take the real energy surface $\Sigma_E$, i.e., $\mathcal{M}_0^* = \Sigma_E \times \{0\}$.

(ii) \emph{Finite-period orbits}: For periodic orbits with a finite $T$, each orbit in real phase space represents a closed trajectory encircling a specific loop \(C\) on the energy surface, parametrized by \((q(t), p(t))\). Upon complexification, \(q(t) \to q(z)\) and \(p(t) \to p(z)\) continue analytically to single-valued functions on a compact Riemann surface \(\mathcal{C}\), around which periodic orbits wrap.

As an illustrative example, consider the one-dimensional double-well potential with the Hamiltonian $H(p, q) = p^2 + q^4 - 2 q^2$. The corresponding solutions are elliptic functions with double periods \(\omega_1\) and \(\omega_2\). Thus, the complexified orbits correspond to cycles on tori, i.e., Riemann surfaces with genus \(g = 1\) (see Fig.~\ref{fig:complex}). For \(E < 0\), one period \(\omega_1\) is real, corresponding to oscillations within a single well, while the other period \(\omega_2\) is purely imaginary, associated with tunneling between wells. For \(E > 0\), the period \(\omega_1\) becomes complex, while \(\omega_2\) remains imaginary. The real periodic orbit corresponds to the cycle \(\omega_1 - 2\omega_2\), reflecting classically allowed motion over the potential barrier. The homology group \(H_1(\mathcal{C}, \mathbb{Z})\) consists of cycles \(n \omega_1 + m \omega_2\) for \(n, m \in \mathbb{Z}\), many of which have complex periods that do not appear in either the standard real-time or imaginary-time path integral formulations. However, these additional cycles contribute non-perturbative effects to the full quantum trace formula.

In general, for a Riemann surface \(\mathcal{C}\) of genus \(g\), the homology group \(H_1(\mathcal{C}, \mathbb{Z})\) is a free abelian group of rank \(2g\), containing \(2g\) independent 1-cycles. Orbits can encircle the handles of the surface in various combinations, corresponding to different homology classes. The reduced action for classical orbits, \(\mathcal{S}[\gamma] = \oint_\gamma p \cdot dq\), has the same value for any cycle \(\gamma\) within the same class because \(p \cdot dq\) is holomorphic and therefore a closed differential on \(\mathcal{C}\). For example, for the double-well potential with \(E < 0\), shifting the variable \(z\) by a complex constant \(c\), the orbit \(q(t + c)\) for real-time \(t\) remains in the same homology class, whereas the imaginary-time orbit \(q(it + c)\) belongs to a different class. The corresponding actions \(\mathcal{S}\) are invariant under such shifts, indicating that these orbits form a continuous family with the same action.

Therefore, each homology class \([\gamma]\) corresponds not to an isolated point but to a critical manifold $\mathcal{M}_\gamma$ of complex dimension one. This additional complex degree of freedom can be decomposed into the direction along the orbit, which corresponds to time translation symmetry and contributes a measure factor proportional to \(|T|\), and a transverse direction that varies across different cycles within the same class. Therefore, to make the middle dimensional, we pick one of the cycles $\gamma \in [\gamma]$ to be the critical manifold $\mathcal{M}_\gamma^*$.

Analogous to the semiclassical trace formula, where real periodic orbits are decomposed into multiples of primitive periodic orbits, homology classes \(H_1(\mathcal{C}_\alpha, \mathbb{Z})\) can similarly be decomposed into multiples of primitive homology classes \(H_1(\mathcal{C}_\alpha, \mathbb{Z})^\mathrm{prim}\), which explicitly separates out repetitive cycles. For a primitive homology class \([p] \in H_1(\mathcal{C}_\alpha, \mathbb{Z})^\mathrm{prim}\), we denote \([rp]\) as its \(r\)-fold multiplicity, and \(\mathcal{M}_{rp}\) as the critical manifold generated by the class \([rp]\). This decomposition is particularly useful when discussing hyperbolic Riemann surfaces with \(g > 1\), where the uniformization theorem allows us to represent \(\mathcal{C}\) as the quotient \(\mathbb{H}/\Gamma\) of the upper half-plane \(\mathbb{H}\) by a Fuchsian group \(\Gamma\). In this context, primitive homology classes correspond to free homotopy classes associated with unique closed geodesics on \(\mathbb{H}/\Gamma\). Thus, integration over a homology class is lifted to integration along the corresponding geodesic in the hyperbolic plane.

\mysec{Quantum trace formula} 
To explore the flow equation of thimbles, we choose the flat metric on the free loop space $|\delta \tilde z(\eta)| = \oint \delta \tilde z^i(\eta) \overline{\delta \tilde z^j(\eta)} d\eta$. The corresponding flow equation is then given by
\begin{subequations}\label{eq:flow}
   \begin{align}
      &\frac{\partial \tilde z_i}{\partial \tau}  = i \overline{\left(\Omega \frac{\partial \tilde z_i}{\partial \eta}  - T \nabla_{i}  H\right)}, \label{eq:flowa}\\
      &\frac{d T}{d \tau} = i  \overline{\left(\int_0^1 (E - H) d \eta \right)} \label{eq:flowb}.
   \end{align}
\end{subequations}
Using real coordinates $y_i \ldef \{\Re \tilde q_i, \Im \tilde p_i, \Im \tilde q_i, \Re \tilde p_i\}$, Equation \eqref{eq:flowa} reduces to the Floer flow equation
\begin{equation}\label{eq:flow1}
\frac{\partial y_i}{\partial \tau} = -J \frac{\partial y_i}{\partial \eta} - \nabla_{i} \Im(TH),
\end{equation}
where the $4\times 4$ matrix $J$ consists of a direct sum of two \( 2 \times 2 \) blocks of the form 
\(
\begin{pmatrix}
0 & -1 \\
1 & 0
\end{pmatrix}
\). Thus, $J^2 = -id$ defines a complex structure. Note that this complex structure differs from the one in the original complexified ambient space. Therefore, this equation can be viewed as a perturbed Cauchy-Riemann equation. The periodic $T$ is a function of $\tau$, adding an additional layer of complexity compared to the standard Floer flow equation. 

For the thimble $\mathcal{J}_0$ attached to the critical manifold $\mathcal{M}_0^*$. The main contribution to the thimble integral $\tilde \Gamma(E) \ldef \int_{\mathcal{J}_0} d \mu e^{i\tilde{\mathcal{S}}[\tilde z]}$ comes from the periodic orbits with small period $T$. In these cases, the flow equation \eqref{eq:flow1} can be approximated by the Cauchy-Riemann equation. By introducing $u \ldef \Re \tilde q + i \Im \tilde p $ and $v \ldef \Im \tilde q + i \Re \tilde p$, the solution of the approximated flow equation consists of holomorphic maps \( f(w) \ldef (u(w), v(w)) \) from the unit disc to $\widehat{M}$, where $w \ldef \exp(2\pi (\tau + i \eta))$ for $\tau \in (-\infty, 0)$. Here, we require the boundary condition \( f(0) = (\Re \tilde q_0 + i \Im \tilde p_0, \Im \tilde q_0 + i \Re \tilde p_0)\), where $(\tilde q_0, \tilde p_0) \in \Sigma_E$. The thimble integration can be factorized into the integration over the zero mode of $f(w)$, which equals the volume of the energy surface $\Gamma(E)$, and higher modes, which lead to a normalization factor independent of $E$. Thus, we have $\tilde \Gamma(E) \approx \Gamma(E)$, aligning with the short-time contributions of the semiclassical formula but with additional quantum corrections from short but non-zero $T$, whose values, however, can only be determined by the full thimble integration.

We are now ready to apply Eq. \eqref{eq:Lefschetz} to the integral \eqref{eq:d}, yielding the full quantum trace formula
\begin{align}\label{eq:qtrace}
d(E) = \tilde \Gamma(E) +  \sum_\alpha \sum_{[p] \in H_1(\mathcal{C}_\alpha, \mathbb{Z})^\mathrm{prim}} \sum_{r=1}^\infty \n_{rp} \mathcal{A}_{rp} \, e^{ir \mathcal{S}_p}.
\end{align}
Here, $\mathcal{A}_{rp} \ldef \frac{1}{2\pi}\int_{\mathcal{J}_{rp}} \d \mu_{\mathcal{J}_{rp}} \,  e^{i{\Delta \tilde{\mathcal{S}}[\tilde z]}{}}$, and $\n_{rp} $ is the interaction number of the dual thimble $\mathcal{K}_{rp}$  with the original real functional space contour. For $\Im \mathcal{S}_p \leq 0 $, $\n_{rp} = 0$ except for the real period orbits, where $\n_{rp}=1$. For $\Im \mathcal{S}_p > 0$, $\n_{rp}$ depends on the flow equation.  

\mysec{Discussion} 
The full quantum trace formula \eqref{eq:qtrace} has a similar structure to the semiclassical trace formula \eqref{eq:sctrace}. In particular, the Maslov index term arises naturally from the flow equation, as shown in \cite{tanizaki2014real}. However, one of the major discrepancies between these two is the contribution from orbits with complex periods, which contributes to the nonperturbative density of states \( e^{-\Im (\mathcal{S}_p)} \). Does the quantum trace formula approximate the semiclassical trace formula by considering only the real periodic orbits? The short answer is no. This is because the Lefschetz thimble construction restricts the flow to directions with positive eigenvalues, thereby limiting contributions to stable orbits by its very construction. In contrast, the amplitude determinant in the semiclassical trace formula includes contributions from both stable and unstable orbits, with the latter playing a significant role in the density of states. Nevertheless, the semiclassical formula offers only an approximation, as it lacks critical nonperturbative information.

Many questions remain open. For instance, determining the intersection number is a notably difficult problem when applying the Lefschetz thimble method. The coupled flow equations pose additional challenges for performing thimble integration. Nevertheless, the thimble integration can be treated perturbatively and is believed to be Borel summable. Finally, as with all trace formulas, enumerating all periodic orbits is a very difficult task, except in limited simple cases. However, it might be possible to bound the energy gap using only the shortest periodic orbits, as was done previously in the Selberg trace formula. This could provide a new tool to examine the mass gap problem in Yang-Mills theory. Of course, this also requires generalizing the present approach to quantum field theory (QFT), where periodic orbits are replaced by periodic instantons—a task we leave for future investigation.

In conclusion, we present a new attempt to derive an exact trace formula by accounting for the contributions of all complex periodic orbits. This naturally accounts for all nonperturbative effects. As analytical and numerical tools for the Lefschetz thimble method rapidly improve \cite{alexandru2022complex}, our trace formula holds the promise of becoming an important means to understand the nonperturbative nature of quantum systems.


\bibliography{ref}

\end{document}